\documentclass[journal=jacsat,manuscript=article]{achemso}
\setkeys{acs}{articletitle=true}

\usepackage{upgreek,textgreek}
\usepackage[version=3]{mhchem} 
\usepackage[T1]{fontenc}       
\usepackage{float}
\usepackage{graphicx}
\usepackage{subfig}
\usepackage{epsfig}
\usepackage{epstopdf}
\usepackage{enumerate}
\usepackage{physics}
\usepackage{longtable}
\usepackage{amssymb}
\usepackage{color}
\usepackage{colortbl}

\definecolor{hellgrau}{rgb}{0.90,0.90,0.90}

\newcommand{\RNum}[1]{\uppercase\expandafter{\romannumeral #1\relax}}
\newcommand{\rNum}[1]{\lowercase\expandafter{\romannumeral #1\relax}}

\author{Davinder Singh}
\email{davinder.mand@utoronto.ca}
\affiliation{Center for Quantum Information and Quantum Control and Chemical Physics Theory Group, Department of Chemistry, University of Toronto, Toronto, Ontario M5S 3H6, Canada}
\author{Chern Chuang}
\email{chern.chuang@unlv.edu}
\affiliation{Department of Chemistry and Biochemistry, University of Nevada, Las Vegas, Nevada, USA}
\author{Paul Brumer}
\email{paul.brumer@utoronto.ca}
\affiliation{Center for Quantum Information and Quantum Control and Chemical Physics Theory Group, Department of Chemistry, University of Toronto, Toronto, Ontario M5S 3H6, Canada}

\title{{\Large Machine Learning Optimization of non-Kasha Behavior and of Transient Dynamics in Model Retinal Isomerization}}


\begin{document}

\begin{tocentry}
{\centering
\includegraphics[width = 1.0 in]{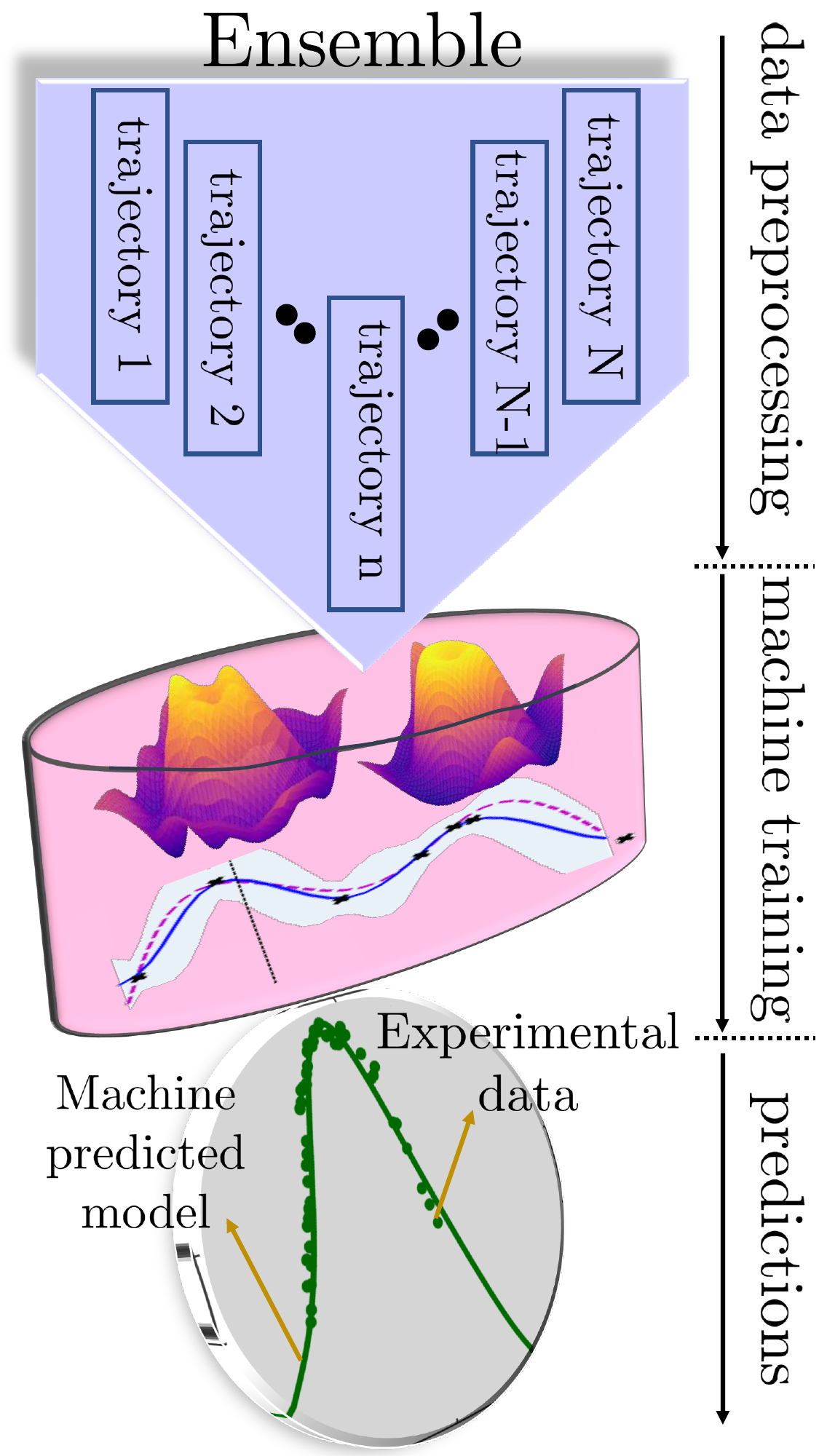}
\par
}
\end{tocentry}

\begin{abstract}
Designing a model of retinal isomerization in Rhodopsin, the first step in vision, that accounts for both experimental transient and stationary state observables is challenging.
Here, multi-objective Bayesian optimization is employed to refine the parameters of a minimal two-state-two-mode ($TM$) model describing the photoisomerization of retinal in Rhodopsin. 
With an appropriate selection of objectives, the optimized retinal model predicts excitation wavelength-dependent fluorescence spectra that closely align with experimentally observed non-Kasha behavior in the non-equilibrium steady state.
Further, adjustments to the potential energy surface within the $TM$ model reduce the discrepancies across the time domain.
Overall, agreement with experimental data is excellent.


\end{abstract}



$Introduction-$Understanding and modeling the isomerization of retinal in Rhodopsin, the first step in human vision, is an important problem that has been the subject of attention for decades\cite{wald_molecular_1968,hurley_temperature_1977,wang_vibrationally_1994,seidner_nonperturbative_1995,hahn_quantum-mechanical_2000,hahn_femtosecond_2000,tscherbul_excitation_2014,
brumer_shedding_2018,johnson_primary_2017,chuang_extreme_2021}. Its complexity presents a number of computational challenges.
Usually, by utilizing structural information, electronic structure calculations are successfully used to estimate potential energy surfaces (PESs) of polyatomic molecules in photobiology, such as the photosynthetic light harvesting Fenna-Matthews-Olson complex\cite{olbrich_atomistic_2011} or the phycobiliprotein light-harvesting complex phycocyanin 645 (PC645) of the cryptophyte algae $Chroomonas$ $mesostigmatica$\cite{mirkovic_ultrafast_2007}.
Within the Born-Oppenheimer approximation, a relatively small error accompanies the estimates of PESs. However, significant error accumulates when two PESs become degenerate at a singular point, leading to a Conical intersection (CI). In the vicinity of a CI, large non-adiabatic coupling between electronic states results in strongly mixed quantum dynamics of the electron and the nucleus that leads breakdown of the Born-Oppenheimer approximation\cite{domcke_role_2012}.
The strong mixing of nuclear and electronic motion hinders straightforward interpretation of model parameters and consequently of the essential physical mechanisms at play. This is especially important in the context of light-induced processes in biological systems including relaxation processes in the DNA\cite{satzger_primary_2006}, internal conversion in $\beta$-carotene\cite{liebel_vibrationally_2014}, and 
the biological vision process \cite{hahn_femtosecond_2000,tscherbul_excitation_2014,
brumer_shedding_2018,johnson_primary_2017,chuang_extreme_2021}.
In these cases specific functionalities may be controlled and moderated by environmental fluctuations and dissipations, and parameter-signal correspondence may be skewed and obscured.
To overcome this challenge, we develop a machine learning based Bayesian optimization method to more accurately determine Hamiltonian parameters, as described below, with a focus on the photoisomerization reaction of Rhodopsin.

Note that even with limited access to the realistic PESs, models of biological light-induced vision processes have been parameterized by restricting the observables to the transient regime, such as those measured using pulsed laser excitation \cite{seidner_nonperturbative_1995,hahn_quantum-mechanical_2000,hahn_femtosecond_2000,johnson_primary_2017}, where the correspondence between parameters and signals is most salient and straightforward.
However, the {\it in-situ} interaction with incoherent (solar) radiation, the natural working condition for these biological systems, requires consideration of the non-equilibrium steady state (NESS) \cite{tscherbul_excitation_2014,dodin_coherent_2016,brumer_shedding_2018}, 
where the correspondence between parameters and signals is highly correlated and convoluted. For example, the NESS quantum yield has been observed to be highly sensitive to the minute change in system parameters, whereas such extreme sensitivity has not been observed in the transient domain \cite{chuang_extreme_2021}. In addition, optimization of one or few observables at the NESS often comes at the cost of sacrificing others.
For example, the NESS emissions spectra of bovine Rhodopsin (measured at three different excitation wavelengths\cite{kochendoerfer_spontaneous_1996}) are poorly described using all existing retinal models. Such a deviation from experimental data is most significant when considering the spectral shift of the emission peak in response to the shift of excitation wavelength, a unique signature of the retinal system that shows disagreement with Kasha's rule. Specifically, with redshifting excitation wavelength, there is a corresponding redshift of the emission spectral peak, whereas Kasha's rule predicts \cite{DF9500900014} that the emission spectra are unchanged with respect to the excitation profile. While good agreement of one emission spectrum (with its peak shifted in accordance with the experiment) is achievable, it has not been shown possible to simultaneously fit all existing data sets \cite{vargas-hernandez_multi-objective_2021}, the challenge addressed in this paper.

To refine the parameters for observables measured at the NESS, our group has recently employed a multi-objective (MO) Bayesian optimization scheme \cite{gopakumar_multi-objective_2018,solomou_multi-objective_2018,chen_machine_2020,hase_chimera_2018,koledina_multi-objective_2019,stevenson_genetic_2019,abranches_stochastic_2024} to determine the extent to which proposed theoretical models\cite{seidner_microscopic_1994,seidner_nonperturbative_1995,hahn_quantum-mechanical_2000,hahn_femtosecond_2000,gonzalez-luque_computational_2000,marsili_two-state_2019} of retinal photoisomerization can be improved \cite{vargas-hernandez_multi-objective_2021}.
This involved optimizing multiple objectives for single trajectories simulated in the NESS using three theoretical models, i.e., a two-state one-mode, a two-state two-mode (refereed as $TM$ model), and a two-state three-mode \cite{vargas-hernandez_multi-objective_2021}.
Various search space parameters were utilized in the optimization process for different retinal models, while the objective space focused on exploring the sum squared error for emission spectra.
However, only the $TM$ model exhibited a separation of emission peaks as a function of excitation frequency, a hallmark of experimentally observed non-Kasha behavior. The lack of quantitative agreement with experimental data remained a challenge. 

In this work, we dramatically improve the $TM$ model via a modified MO Bayesian optimization algorithm which explores an updated search and objective space by employing a different state-of-the-art acquisition function and optimizer using BoTorch \cite{daulton_differentiable_2020,daulton_parallel_2021}. 
In addition, we optimize beyond a single trajectory (described by a single set of parameters). That is, to the best of our knowledge, all existing measurements of emission spectra of Rhodopsin involve ensemble averages over a macroscopic number of molecules.
Therefore, to mimic such effects, we average over an ensemble of fixed width with a normally distributed optical gap between the two electronic states. By doing so, the inhomogeneity of the ensemble is leveraged, and each of the samples gives rise to its own particular response to a given excitation wavelength.
As shown below, these modifications lead to much improved emission spectra at the NESS, such that the updated objective space results in accurate peak positions with acceptable line-shape. However, challenges remained in accurately predicting photoproduct rise time under pulsed laser excitation, an observable belonging to an increasingly available set of experimental data \cite{johnson_primary_2017}. 
This implied that even with the inclusion of ensemble averaging, a minimal $TM$ model is not sufficient to explain both the transient and the stationary state aspects simultaneously. 
Accordingly, as discussed below, we modify the PES by incorporating terms corresponding to finer details of the reaction coordinate, notably improving the theoretical predictions across the short and long time domains.
\textit{Overall, this result provides a microscopic model for retinal isomerization in Rhodopsin based on the interplay between the dissipation of electronic energy via coupling to fast phonon degrees of freedom and the variation of optical gap afforded by slow motions of the protein environment on time scales longer than the establishment of the NESS under the open quantum system framework.} Details are provided below. Note at the outset, however, that our approach can readily apply to a wide variety of systems. 




$Minimal$ $TM$ $model-$The $TM$ model of retinal isomerization in Rhodopsin is discussed in detail in Ref[\cite{chuang_extreme_2021,chuang_steady_2022,vargas-hernandez_multi-objective_2021}]. The observed experimental deviation from Kasha's rule \cite{kochendoerfer_spontaneous_1996} indicates that the relaxation processes in retinal are on the same time scale or faster than the emissive decay from the excited state. Such fast relaxation is accounted for in the minimal $TM$ model by including a conical intersection between two electronic states and a strongly coupled harmonic mode embedded in a dissipative environment. 
The system Hamiltonian composed of two diabatic electronic states, $\ket{0}$ and $\ket{1}$, can be written as \cite{hahn_quantum-mechanical_2000,hahn_femtosecond_2000}

\begin{align}
\hat{H}_s &= \hat{T} \begin{bmatrix} 1 & 0 \\
0 & 1 
 \end{bmatrix} + \underbrace{\begin{bmatrix} V_{00} & V_{10} \simeq \lambda x \\
V_{01} \simeq \lambda x & V_{11} 
 \end{bmatrix}}_{\mathcal{M}},
\label{ham0}
\end{align}
where $\hat{T} = -\left( \frac{1}{2m} \right) \frac{\partial^2}{\partial \phi^2} + \frac{\omega}{2} \frac{\partial^2}{\partial x^2}$ describes the kinetic energy operator with associated effective mass $m$.
The inverse mass parameter $m^{-1} = 22.6$ cm$^{-1}$ and $\omega = 1532.4$ cm$^{-1}$ are held fixed in the optimization procedure.
Here $V_{10} = V_{01} \simeq \lambda x$, and the diabatic potential energy surfaces $V_{jj}$ (PES) for the ground ($j=0$) and excited ($j = 1$) states are expressed as
\begin{align}
V_{00} &\simeq \frac{1}{2}W_0 \left[ 1 - \cos\left( \phi \right) \right] + \frac{\omega}{2} x^2 \nonumber\\
V_{11} &\simeq E_1 - \frac{1}{2}W_1 \left[ 1 - \cos\left( \phi \right) \right] + \frac{\omega}{2} x^2 + \kappa x\;,
\label{v00}
\end{align}
where $E_1$ represents the energy gap between the two electronic states. $\mathcal{M}$ indicates the potential energy part of the system Hamiltonian. By fixing the energy storage $\Delta E$ (i.e., the energy difference between the {\it cis} and {\it trans} retinal conformers) to its experimental value, $W_1$ can be estimated using $\Delta E(=E_1 - W_1)$ with the assumption that the electronic gap between excited and ground states match the center frequency of the absorption bands of the two isomers \cite{hahn_quantum-mechanical_2000}. Similarly, $W_0$ can be related to the frequency along the isomerization coordinate $\phi$ \cite{hahn_quantum-mechanical_2000}. An additional vibrational degree of freedom is included in PES via $x$, with a shift in its equilibrium position between the two electronic states indicated by $\kappa$.
In addition, the system-bath interaction and relevant parameters are summarized in the Supporting Information (SI). 
Note that reasonably accurate values of the majority of parameters used to describe the PES such that optical gap $E_1$, $W_0$, $W_1$, $\lambda$, and $\kappa$ are largely inaccessible in experiments directly. 
Moreover, the non-trivial processes to find these parameters are computationally costly.
To overcome this challenge, we employ machine-learning based surrogate models\cite{snoek_practical_2012,daulton_differentiable_2020,daulton_parallel_2021} to refine the parameters in the optimization by comparing the theoretically predicted objectives to the corresponding experimental observables.
To do so, by setting $E_1 - W_1 =11211$ cm$^{-1}$($= 1.39$ $eV$\cite{schick_energy_1987}), the search space parameters in the optimization of the $TM$ model are $\Theta = \left[ E_1+\delta E_1, W_0 + \delta W_0,\kappa + \delta \kappa, \lambda + \delta \lambda \right]$ where dimension of search space is $d=4$. 
For simplicity, each search space point can be expressed as $\theta = \{\theta_d^0\} + \{\delta \theta_d\}$, with $\{\delta\theta_d = 0\}$ referring to the parameters\cite{johnson_primary_2017} listed in Table \ref{table_TMS}. 
Next, in the MO Bayesian optimization (see the SI), an acquisition function is used to iteratively select new points by varying the values of $\{\delta\theta_d\}$ with bounds such that $-7662 \leq \delta\theta_1 = \delta E_1 \leq 16000$, $-12502 \leq \delta\theta_2 = \delta W_0 \leq 7662$, $-7662 \leq \delta\theta_3 = \delta\kappa, \delta\theta_4 = \delta\lambda \leq 7662$ (all in cm$^{-1}$). By considering $256$ quasi-MC samples, the acquisition function uses all parallel algorithms with sequential greedy optimization (for selecting a batch of new data points). 
For optimization, L-BFGS-B is used to optimize the acquisition function while $\mathcal{GP}$ models are optimized with TNC.
To restart, 20 starting points for multistart acquisition function optimization are used with 1024 samples for initialization.
Further, to include the effect of the disordered ensemble on the emission spectra $g_{th}^{\nu_m}$ measured after the monochromatic excitation at wavelength $\nu_m$ (described in the SI), a normal distribution around $E_1 + \delta E_1$ is used with a full width at half maximum of $\Delta E_1^{w}$. 
In the objective (or target) space, the following error function is adopted $f_{\nu_m}(\theta) = \mid \nu_{ex}^{peak} - \nu_{th}^{peak} \mid$,
where $\nu_{ex}^{peak}$ ($\nu_{th}^{peak}$) represents experimentally observed (theoretically predicted) frequency corresponding to the peak of the emission spectrum excited at $\nu_m$, i.e., $g_{ex}^{\nu_m}(\theta)$ ($g_{th}^{\nu_m}(\theta)$). We consider three emission spectra\cite{birge_nature_1988} (excited at $\nu_1 = 20834$ cm$^{-1}$, $\nu_2 = 19334$ cm$^{-1}$ and $\nu_3 = 17667$ cm$^{-1}$); hence the dimension of the objective space $M=3$.

\begin{figure}[!h]
\begin{center}
\includegraphics[width = 6.5 in]{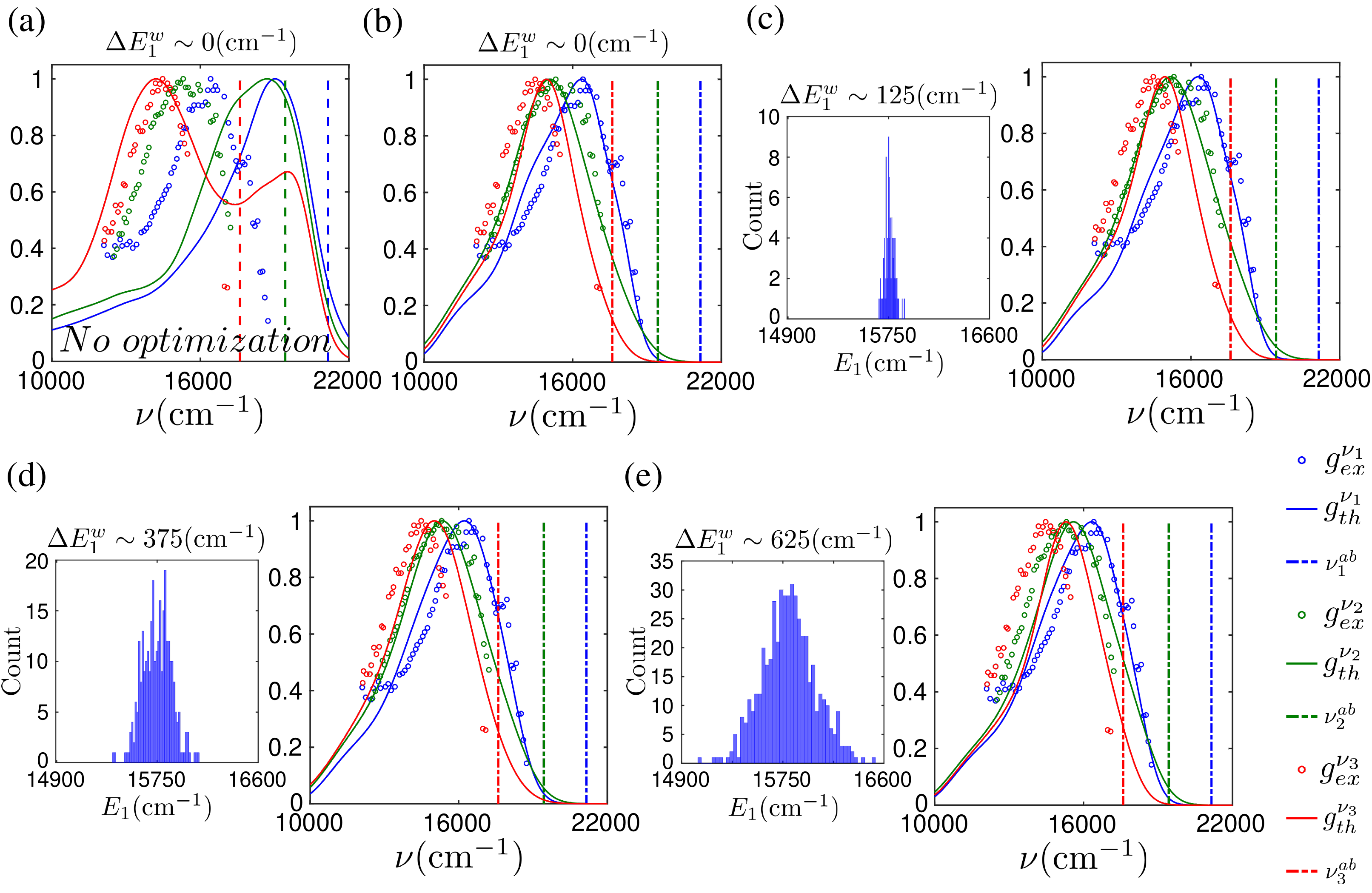}
\caption{ (a) The emission spectra predicted by $TM$ model with distribution width of $ 0$ cm$^{-1}$ without optimization. The emission spectra after Bayesian optimization with distribution width (b) $ 0$ cm$^{-1}$ (c) $\sim 125$ cm$^{-1}$ (d) $\sim 375$ cm$^{-1}$ and (e) $\sim 625$ cm$^{-1}$. Histograms are shown for the corresponding distribution width. The experimental observations [$g_{ex}^{\nu_m}$ where $m = 1,2,3$] are shown by circles and $g_{th}^{\nu_m}$ describes the theoretically calculated spectra. The corresponding absorption frequencies (i.e., $\nu_m^{ab}$) are displayed by vertical lines.}
\label{Emission}
\end{center}
\end{figure}

\begin{table}[H]
  \centering
  \setlength\belowcaptionskip{5 pt}
  \caption{$TM$ model system parameters (in cm$^{-1}$)}
    \begin{tabular}{cccccc}  
  \cellcolor{hellgrau} Optimization  & \cellcolor{hellgrau} $E_1$   &\cellcolor{hellgrau} $W_0$  &\cellcolor{hellgrau} $W_1$  &\cellcolor{hellgrau} $\kappa$ &\cellcolor{hellgrau}  $\lambda$   \\
  No\cite{johnson_primary_2017} ($\{\delta\theta_d = 0\}$)  & 20809.0     & 28713.0  &  9598.0    &   1532.5   & 1532.4     \\
  Yes ($\theta^b$)  & 15764.8     & 23289.0  &  4553.8    &   2080.1   & 2500.3     \\
    \end{tabular} 
    \vspace{0.6 \baselineskip} \\
    \raggedright
  \label{table_TMS}%
  \vspace{-1.8 \baselineskip}
\end{table}
Multiparameter optimization requires that we find the best Pareto frontier $\mathcal{PF}$ (described in the SI). To do so, four different acquisition functions along with the random search are compared.
Here, the best point $\theta^b$ (tabulated in Table \ref{table_TMS} and shown in the SI) in the $\mathcal{PF}$ for each optical gap distribution width $\Delta E_1^{w}$ covered is presented in Fig.~\ref{Emission}.  For comparison, the single trajectory result corresponding to the pre-optimized $TM$ model, ($\{\delta\theta_d = 0\}$) is shown in Fig.\ref{Emission}a. With optimization, $[\delta E_1 = -5044.2, \delta W_0 = -5424.0, \delta \kappa = 547.6, \delta \lambda = 967.9]$
, even for $\Delta E_1^{w} \sim 0$ cm$^{-1}$, the predicted emission spectra $g_{th}^{\nu_1}$ (blue solid line) and $g_{th}^{\nu_2}$ (green solid line) show significant improvement in terms of the peak position and the line-shape, while for $g_{th}^{\nu_3}$ (red solid line) the spectral line-shape is greatly improved.  Overall, we find a reasonably small peak shift ($\sim 650$ cm$^{-1}$) of $g_{th}^{\nu_3}$ from the target in this case [Fig.\ref{Emission}b]. Comparative analysis of Figs.\ref{Emission}b, c and d confirms that an intermediate value of $\Delta E_1^{w}$ gives the best agreement with the experimental spectra. For instance, $g_{th}^{\nu_1}$ in the range $\sim 12000$ to $\sim 16000$ cm$^{-1}$ shifts close to experimental data as we increase the ensemble width from $0$ cm$^{-1}$ [Fig.\ref{Emission}b] to $\sim 125$ cm$^{-1}$ [Fig.\ref{Emission}c]. Similar improvement can be found going from $\Delta E_1^{w} \sim 125$ cm$^{-1}$ to  $\sim 375$ cm$^{-1}$. The best emission line-shape is found at $\Delta E_1^{w} \sim 375$ cm$^{-1}$ [see Fig.\ref{Emission}d] with evident separation of $g_{th}^{\nu_2}$ and $g_{th}^{\nu_3}$ in almost the entire frequency domain. 
Note that further increasing the width ($\Delta E_1^{w} > 375$ cm$^{-1}$) leads to a deteriorating result, as shown by Fig.\ref{Emission}e, where all the predicted emission peaks move towards the $g_{ex}^{\nu_1}$. 
The existence of such an optimal ensemble width is discussed in detail in the SI.
\begin{figure}[!h]
\begin{center}
\includegraphics[width = 4.5 in]{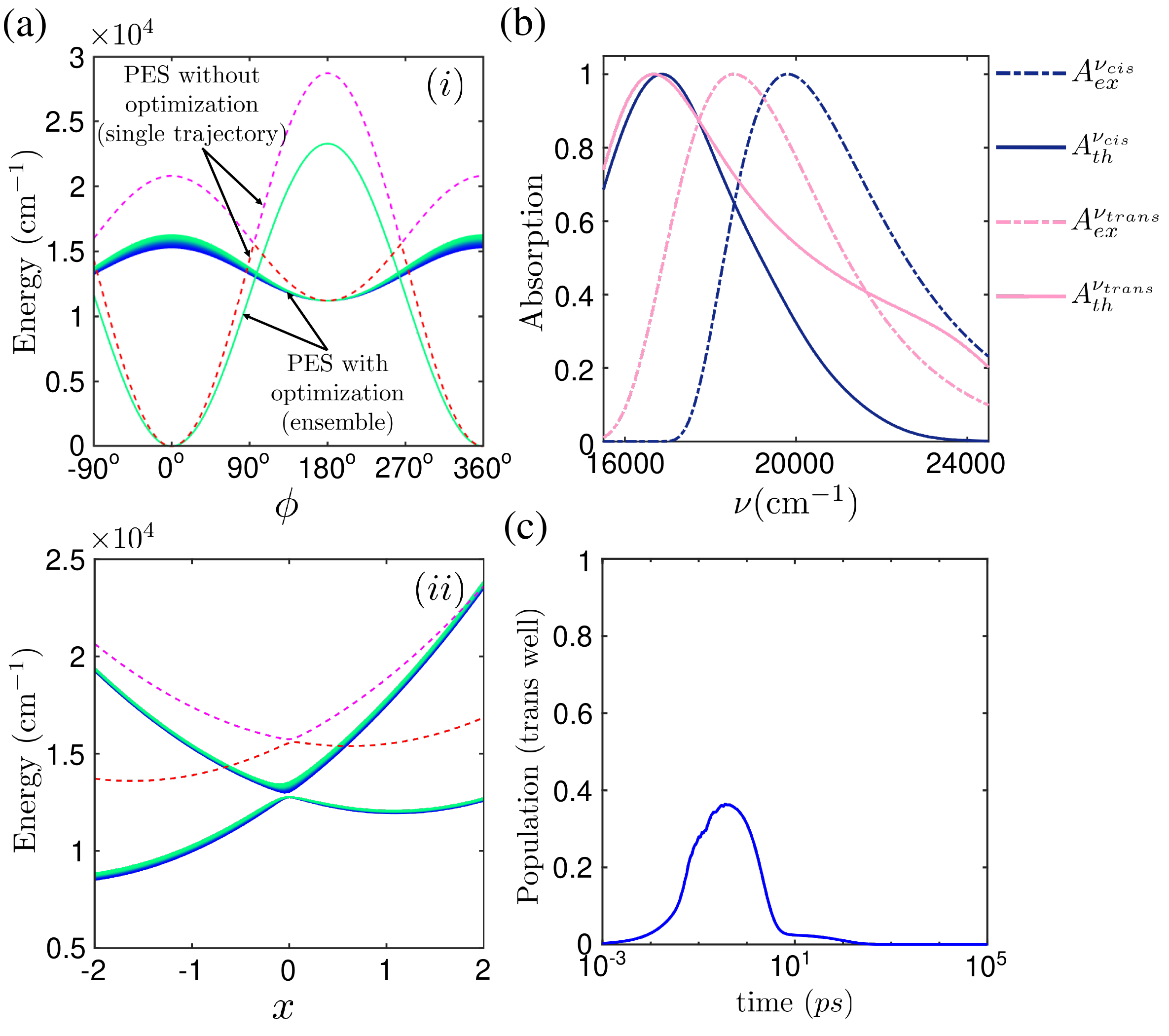}
\caption{ (a) Adiabatic potential energy surfaces ($\rNum{1}$) by varying $\phi$ at $x = 0$ and ($\rNum{2}$) by varying $x$ at $\phi \sim 90^\circ$. Dotted lines illustrates potential energy surfaces without optimization for single trajectory and solid lines shows optimized potential energy surfaces using $\theta^b$ with ensemble width $\Delta E_1^{new} \simeq 375$ cm$^{-1}$.  (b) The absorption spectra and (c) the time-dependence of the $trans$-well population for the $TM$ model with a single trajectory.}
\label{PES}
\end{center}
\end{figure}

Additional insights can be obtained from examining the adiabatic PES. A cut through the PES is shown in Fig.\ref{PES}a. Here we compare the pre-optimized model ($\{\delta \theta_d = 0 \}$, dashed lines) with the best set found by the MO Bayesian optimization ($\theta^b$, solid lines). First, note that the former is from a single set of parameters ($\Delta E_1^{w}=0$) and is therefore represented as a single pair of PES, whereas the latter represents a collection of PES (colored). Though this is the case, it is worth noting that the collection is only narrowly distributed, with the width being $\sim2\%$ of the total optical gap. In the optimized parameter set, the slope of the PES is reduced along the reaction coordinate $\phi$ in the vicinity of the conical intersection around $\phi \sim \pm 90^\circ$ [see Fig.\ref{PES}a($\rNum{1}$)]. This, in conjunction with a significant increase in the non-adiabatic coupling $\lambda$ in the optimized model, leads to modified topography near the conical intersection. This is consequential to the relaxation dynamics, leading to a shift in the emission peak frequency consistent with the experimental observation.

Although the fit to the experimental data is much improved, there are still discrepancies between the absorption spectra simulated in $cis-$ or $trans-$ conformations (described in the SI) and the experimental data, as shown in Fig.\ref{PES}b. These discrepancies were also present in previous attempts using single trajectory models, so incorporating ensemble averaging does not resolve the issue. In comparison to the experimental values, where $\nu_{ex}^{peak} = 19807$ cm$^{-1}$ of $A_{ex}^{\nu_{cis}}$ ($\nu_{ex}^{peak} = 18558$ cm$^{-1}$ of $A_{ex}^{\nu_{trans}}$), the calculated peak frequency of $A_{th}^{\nu_{cis}} $ is $\nu_{th}^{peak} \simeq 16863$ cm$^{-1}$ ($A_{th}^{\nu_{trans}} $ is $\nu_{th}^{peak} \simeq 16667$ cm$^{-1}$). Furthermore, with $\theta^b$, the predicted $trans$ activation energy ($E^{act}_{th}$) is too small, by $159$ cm$^{-1}$.  Consequently, despite a considerable $trans$ well population 
of $\sim 28\%$ at $\sim 101$ $fs$ (compared to the experimental rise time $\sim$ 60 $fs$)\cite{johnson_primary_2017}, the long-time peak quantum yield ($QY$) is negligible, as shown in Fig.\ref{PES}c. 


\textit{Modified} $TM$ $model-$To improve the modeling, we adopt a more comprehensive Bayesian optimization approach. This involves including absorption spectra, $trans$ activation energy, and long-time $QY$ as additional targets in the objective space, while removing the constraint $E_1 - W_1 =\Delta E_{ex}$. Moreover, to simultaneously account for the emission spectra, absorption spectra, long-time $QY$, and $trans$ activation energy, modifications are made to the $TM$ model adiabatic PES as follows:
\begin{align}
V_{00} &= \frac{1}{2}W_0\left[1 + \cos\left(\phi\right) \right] + W_0^a \cos(2\phi) + \frac{\omega}{2}x^2, \nonumber\\
V_{11} &= E_1 - \frac{1}{2}W_1\left[1 + \cos\left(\phi\right) \right] + W_1^a \cos(2\phi) + \frac{\omega}{2}x^2 + \kappa x.
\end{align}
That is, the spatial resolution reaction coordinate is augmented with the next term in the Fourier series, i.e., $\cos(2\phi)$, with the addition of two new search space parameters $W_0^a$ and $W_1^a$. Now, the search space parameters is 
$\Theta = \left[ E_1+\delta E_1, W_0 + \delta W_0, W_0^a + \delta W_0^a,\right.$ $\left.W_1 + \delta W_1, W_1^a + \delta W_1^a,\kappa + \delta \kappa, \lambda + \delta \lambda \right]$ 
with $d=7$. We begin the optimization with $W_0^a = W_1^a = 0$. During the iterative selection 
of the acquisition function, the bounds on additional search space parameters are $-8872 \leq \delta W_1 \leq 16000$, $0 \leq \delta W_0^a \leq 4839$, $0 \leq \delta W_1^a \leq 2420$ (all in cm$^{-1}$). 
Here, instead of $256$, $1024$ quasi-MC samples are used for the acquisition function, and also only L-BFGS-B is used to optimize both the acquisition function and models.
Next, to estimate the error functions in the target space, the theoretically predicted peak frequency $\nu_{th}^{peak}$ corresponding to the absorption spectra $A_{th}^{\nu_{cis(trans)}}$ is compared with the experimentally observed peak frequency $\nu_{ex}^{peak}$ corresponding to the absorption spectra $A_{ex}^{\nu_{cis(trans)}}$ such that $f_{\nu_{cis(trans)}}(\theta) = \mid \nu_{ex}^{peak} - \nu_{th}^{peak} \mid$.
Similarly, the error function accounting for the energy storage, $trans$ activation energy, and long time $QY$ is defined as $f_{\Delta E}(\theta) = \mid \Delta E_{ex} - \Delta E_{th}(\theta) \mid$, $f_{\Delta E^{act}}(\theta) = \quad\mid E_{cns}^{act} - E_{th}^{act}(\theta) \mid$, $f_{\delta QY^{lt}}(\theta) = \mid QY_{ex}^{lt} - QY_{th}^{lt}(\theta) \mid$,
where\cite{schick_energy_1987} $\Delta E_{ex}=11211 \quad \text{cm}^{-1}$ ($\Delta E_{th}$) illustrates the experimentally observed (theoretically predicted) energy storage. Moreover, a target $trans$ activation energy $ E_{cns}^{act} =3226 \quad \text{cm}^{-1}$ is used following the parameters provided in the $TM$ model in Ref[\cite{johnson_primary_2017}] and is compared with theoretically predicted $trans$ activation energy $\Delta E_{th}^{act}$. 
Similarly, the experimental value\cite{kim_wavelength_2001} of the long time $QY$ ($QY_{ex}^{lt}$), that is, 65\% is compared with theoretical predictions $QY_{th}^{lt}$. 
These modifications lead to a much improved Pareto point (referred to as $\theta^b_{add}$ and shown in the SI) with the parameters tabulated in Table \ref{table_TMS_new}.
\begin{table}[H]
  \centering
  \setlength\belowcaptionskip{5 pt}
  \caption{Optimized $TM$ model system parameters $\theta^b_{add}$ (in cm$^{-1}$)}
    \begin{tabular}{ccccccc}  
   \cellcolor{hellgrau}  \cellcolor{hellgrau} $E_1$   &\cellcolor{hellgrau} $W_0$ &\cellcolor{hellgrau} $W_0^a$  &\cellcolor{hellgrau} $W_1$ &\cellcolor{hellgrau} $W_1^a$  &\cellcolor{hellgrau} $\kappa$ &\cellcolor{hellgrau}  $\lambda$    \\
    20809.0     & 29828.8 & 3540.7 &  11474.8 & 1405.8 &   -2500.3   & 49.2     \\
    \end{tabular} 
    \vspace{0.6 \baselineskip} \\
    \raggedright
  \label{table_TMS_new}%
  \vspace{-1.8 \baselineskip}
\end{table}
\begin{figure}[!h]
\begin{center}
\includegraphics[width = 6.5 in]{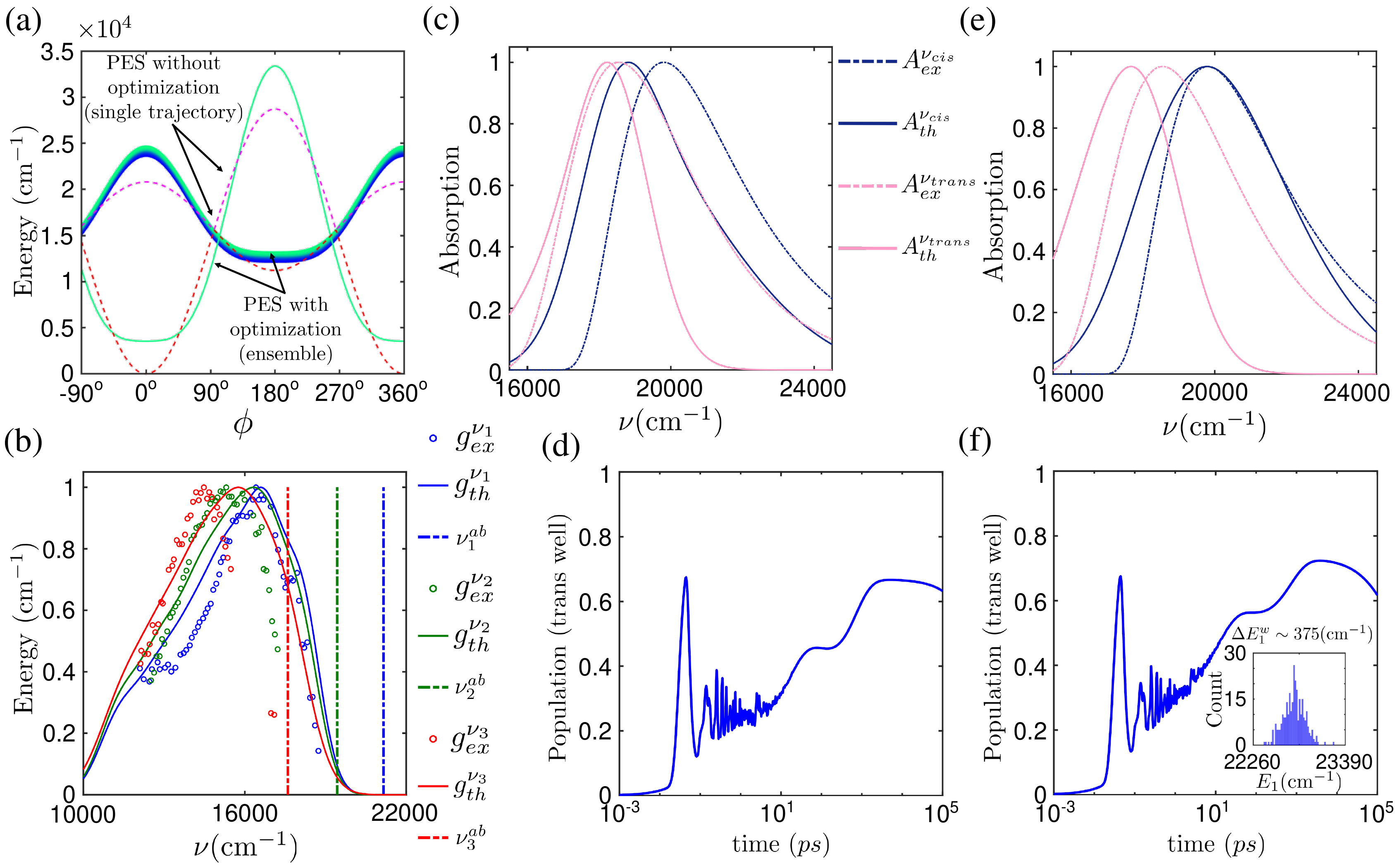}
\caption{ (a) The adiabatic potential energy surfaces of $\theta^b_{add}$ as functions of $\phi$ at $x = 0$. Dotted lines illustrate the pre-optimized PES and solid lines show the optimized PESs using $\theta^b_{add}$ with ensemble width $\Delta E_1^{w} \simeq 375$ cm$^{-1}$. (b) The emission spectra, (c) single trajectory absorption spectra, and (d) dynamics of photoproduct population for single trajectory case.  (e) The absorption spectra and (f) dynamics of photoproduct population with distribution width of $\sim 375$ cm$^{-1}$.}
\label{PES_emm_abs}
\end{center}
\end{figure}
The inclusion of $\cos(2\phi)$ terms in $\theta^b_{add}$ has notable effects on the system. It enhances the \textit{trans} activation energy, resulting in $E^{act}_{th} \simeq 6400$ cm$^{-1}$ for the single trajectory case. Additionally, it leads to a flattening of the \textit{trans}-well, as depicted in Fig.\ref{PES_emm_abs}a. These modifications not only improve the theoretically predicted \textit{trans} absorption spectrum for the single trajectory case (Fig.\ref{PES_emm_abs}c) but also significantly enhance the long-time $QY$ (Fig.\ref{PES_emm_abs}d). The $trans$ well population 
in this scenario peaks at around $44$ $fs$. Furthermore, the emission spectrum $g_{th}^{\nu_1}$ shows remarkable accuracy (Fig.\ref{PES_emm_abs}b), while $g_{th}^{\nu_2}$ and $g_{th}^{\nu_3}$ exhibit slight blue shifts compared to experiment.

\begin{figure}[!h]
\begin{center}
\includegraphics[width = 4.5 in]{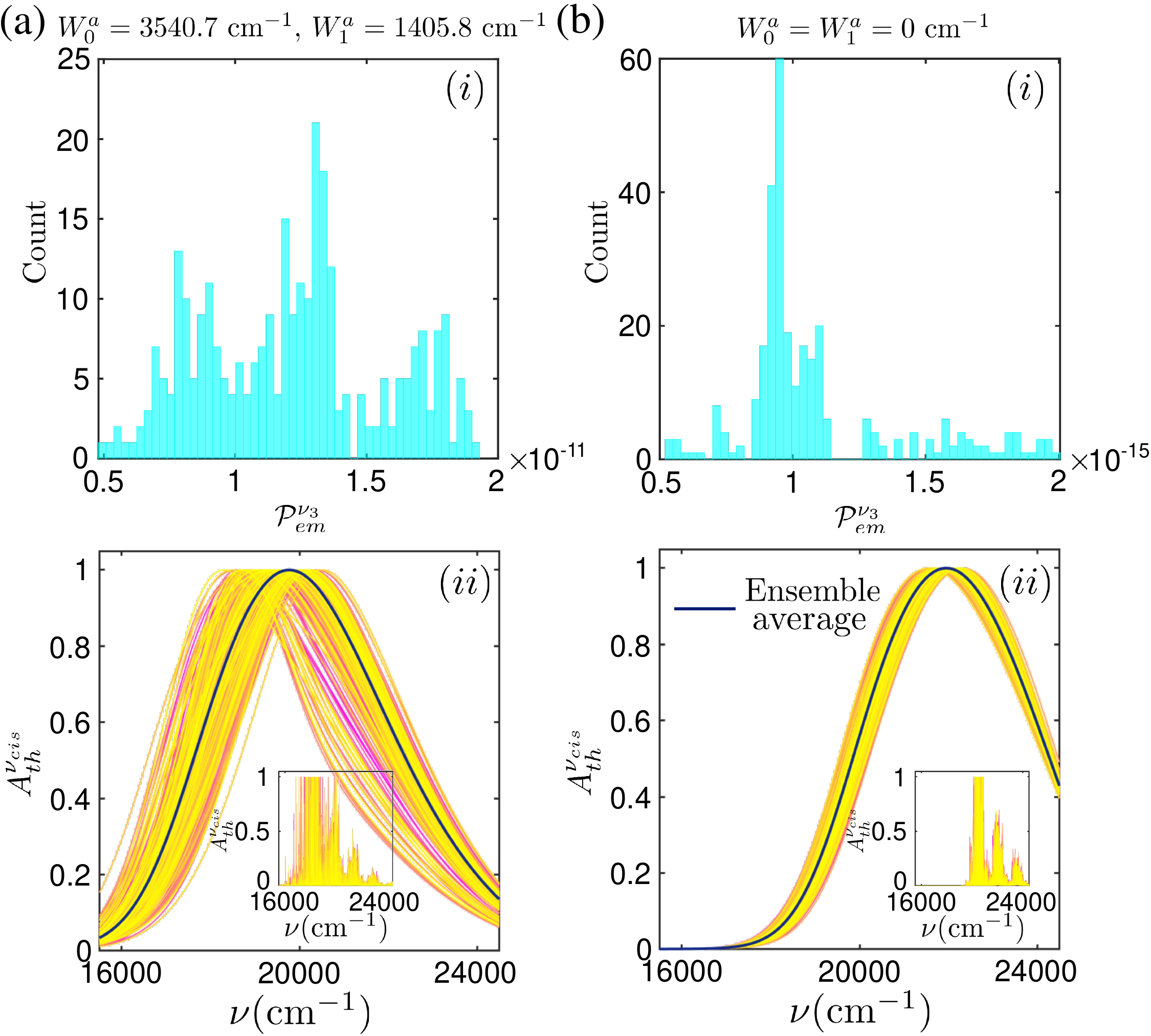}
\caption{ ($\rNum{1}$) Total emissive power $\mathcal{P}^{\nu_3}_{em} \simeq \int d\nu g_{th}^{\nu_3}$ and ($\rNum{2}$) $cis$ absorption spectrum for different trajectories (a) with addition and (b) without addition of $\cos(2\phi)$ term. Inset: Emission spectrum for each trajectory before averaging. Here, we used $\Delta E_1^w \sim 375$ cm$^{-1}$.}
\label{Bar_sum_env_eff}
\end{center}
\end{figure}

In addition to the single trajectory case, we explored ensemble averages centered at the selected Pareto point $\theta^b_{add}$with different disorder strengths. With increasing width, the $cis$ absorption improves (Fig.\ref{PES_emm_abs}e) and the \textit{trans} activation energy decreases initially with ensemble width and plateaus at $E^{act}_{th} \sim 5000$ cm$^{-1}$ after reaching a width $\Delta E_1^w \approx$ 375 cm$^{-1}$. 
The transient behavior up to approximately $10$ ps remains largely unaffected by the ensemble width, and the photoproduct population peaks at 44 fs. Only a slight variation is observed in the temporal dynamics after $\sim 1$ ns (see Fig.\ref{PES_emm_abs}f). Again, the emission peaks are not significantly affected by using an ensemble for $\theta^b_{add}$.
This observation is consistent with our comparative analysis of the contribution to the emission within an ensemble, where we gathered the statistics of the total emissive power ($\mathcal{P}^{\nu_j}_{em} \simeq \int d\nu g_{th}^{\nu_j}$) in each trajectory (the area under the emission spectrum).
The emissive power histograms reveal that in the presence of $\cos(2\phi)$ terms, almost all trajectories in the ensemble contribute rather evenly to the emission [see Fig.\ref{Bar_sum_env_eff}a($\rNum{1}$)].
This equal contribution around the mean value results in a negligible shift of the average emission spectra in the ensemble case. Conversely, in the absence of $\cos(2\phi)$ terms, only a few trajectories made significant contributions [see Fig.\ref{Bar_sum_env_eff}b($\rNum{1}$)], which could impact the mean emission spectra. \textit{This suggests that the inclusion of $\cos(2\phi)$ terms establishes a nearly constant emission quantum yield across the scanned wavelengths, as experimentally observed \cite{kochendoerfer_spontaneous_1996}.} It appears that the addition of $\cos(2\phi)$ terms in the retinal model facilitates a wide distribution of bright states. To confirm this, the $cis$ absorption spectrum is examined, which illustrates that with the inclusion of $\cos(2\phi)$ terms, the absorption peaks are broadly distributed among different trajectories [see Fig.\ref{Bar_sum_env_eff}a($\rNum{2}$)]. Conversely, in the absence of $\cos(2\phi)$ terms, the $cis$ absorption peaks are relatively concentrated for different trajectories [see Fig.\ref{Bar_sum_env_eff}b($\rNum{2}$)]. Examining individual trajectories further reveals that $cis$ absorbs only in the high-frequency domain (i.e., from $\sim 20000$ cm$^{-1}$ to $\sim 24000$ cm$^{-1}$ [see inset Fig.\ref{Bar_sum_env_eff}b($\rNum{2}$)]), whereas the $\cos(2\phi)$ terms broaden the absorption to include the low-frequency range, and $cis$ absorbs from $\sim 16000$ cm$^{-1}$ to $\sim 24000$ cm$^{-1}$ [see inset Fig.\ref{Bar_sum_env_eff}a($\rNum{2}$)]).

$Conclusions-$A multi-objective Bayesian approach has been successfully used to optimize the parameters of a basic model that describes the NESS in the photoisomerization of retinal chromophores in rhodopsin, specifically the {\it cis-trans} isomerization. 
Particular emphasis was placed on the fluorescence emission spectra, which exhibit deviations from the predictions of Kasha's rule and that serve as a distinctive feature of the photoisomerization process. Remarkably, we achieved unprecedented agreement with the experimental emission line-shape by incorporating parameters specified by $\theta^b$ (see Table \ref{table_TMS}) into the model. While a single set of model parameters and a single trajectory yield reasonable agreement for the emission peak position as a function of the excitation frequency, introducing an optimally broadened ensemble of parameter sets enhances the overall line-shape. 

The modified objective space and ensemble averaging improve NESS emission spectra, but it is not entirely satisfactory for other observables like $trans$ well photoproduct rise time in pulsed laser experiments. 
This suggests an inconsistency across scenario domains, where non-optimized parameters perform well for short-time pulsed laser induced behavior but fail to produce satisfactory NESS emission spectra, whereas optimized parameters $\theta^b$ excel in NESS emission spectra but not in short-time results. 
This deficiency motivated modifying the minimal $TM$ model to predict outcomes across both time domains,
by incorporating additional terms in the adiabatic PES.
This modification lead to improved theoretical predictions, \textit{with a single set of optimized parameters $\theta^b_{add}$ [Table \ref{table_TMS_new}] giving reasonable accuracy in both short and long time domains.
Incorporating additional averaging did not undermine the existing fits.} 

These insights and theoretical framework hold promise for application in more complex models and steady-state processes prevalent in photophysics and photochemistry, particularly in scenarios where static disorder or vibrational degrees of freedom with time scales much slower than the relaxation of optical excitation contribute to the observables.
In the case of photoisomerization of retinal in Rhodopsin our results motivate new experiments, required in order to further develop models of the isomerization process.

%
%
%

\begin{suppinfo}

This material is available. 
	
\end{suppinfo}

\begin{acknowledgement}
This material is based upon work supported by the Air Force Office of Scientific Research under award FA9550-20-1-0354.
\end{acknowledgement}

\setcounter{section}{0}
\renewcommand{\thesection}{S\arabic{section}}

\setcounter{figure}{0}
\renewcommand{\thefigure}{S\arabic{figure}}
\setcounter{equation}{0}
\renewcommand{\theequation}{S\arabic{equation}}
\setcounter{table}{0}
\renewcommand{\thetable}{S\arabic{table}}

\newpage
\begin{center}
\underline{{\Large \textbf{Supporting Information:}}}
\end{center}

\section{System bath interaction}

To account for the influence of the vibrational degrees of freedom not included in the $TM$ model, such as other modes in the retinal residue as well as those of the opsin protein pocket, the system-bath interaction is approximated as follows \cite{balzer_modeling_2005,hahn_ultrafast_2002}.
\begin{align}
\hat{H}_{sb} = \sum\limits_b \hat{S}^{(b)} \otimes \hat{B}^{(b)} \;,
\label{sys_bath}
\end{align}
where $\hat{S}^{(b)}$ ($\hat{B}^{(b)}$) is system (bath) operator and the superscript $b$ includes all involved baths. The harmonic approximated bath (i.e., $\hat{H}_{b} = \sum_k \omega_k \hat{b}_k^\dagger \hat{b}_k$) consists of two components:
\begin{align}
\hat{H}_{sb} &= \hat{H}_{sx} + \hat{H}_{s\phi}, \nonumber\\
\hat{H}_{sx} &= \hat{S}_x \otimes \sum\limits_k g_{k,x} \left( \hat{b}_{k,x}^{\dagger} + \hat{b}_{k,x} \right), \nonumber\\
\hat{H}_{s\phi} &= \hat{S}_\phi \otimes \sum\limits_k g_{k,\phi} \left( \hat{b}_{k,\phi}^{\dagger} + \hat{b}_{k,\phi} \right) \;,
\label{sys_bath1}
\end{align}
where $g_{k,x}$ and $g_{k,\phi}$ represent the displacement of $k$th phonon mode, and $b_{k,x(\phi)}^{\dagger}$ ($b_{k,x(\phi)}$) denote the corresponding creation (annihilation) operators.  The system operators are $\hat{S}_x =  \ket{1}\bra{1} \cdot x$ and $\hat{S}_\phi =  \ket{1}\bra{1} \cdot \left[ 1 - \cos(\phi) \right]$ \cite{chuang_extreme_2021,chuang_steady_2022}.  Following Balzer and Stock \cite{balzer_modeling_2005}, the Ohmic spectral density with exponential cutoff, i.e., $J_{x(\phi)} = \eta_{x(\phi)}\omega e^{-\omega/\omega_{c,x(\phi)}}$, is used with parameters listed in Table \ref{table_TMB}. 

\begin{table}[H]
  \centering
  \setlength\belowcaptionskip{5 pt}
  \caption{\textit{TM model bath parameters}}
    \begin{tabular}{cccc}  
     \cellcolor{hellgrau} $\eta_x$ $^a$   &\cellcolor{hellgrau} $\eta_\phi$ $^a$ &\cellcolor{hellgrau} $\omega_{c,x}$  &\cellcolor{hellgrau} $\omega_{c,\phi}$ $^b$ \\
    0.1     & 0.15  &  1532.4 cm$^{-1}$    &   572.6 cm$^{-1}$   \\
    \end{tabular} 
    \vspace{0.6 \baselineskip} \\
    $^a$dimensionless parameters, $^b$ $\omega_{c,\phi} \simeq \sqrt{W_0/2m}$.
  \label{table_TMB}%
  \vspace{-0.9 \baselineskip}
\end{table}

\section{Absorption and emission spectra}
By using the Markovian-Redfield framework and assuming the Bloch-secular approximation, the population of energy eigenstates $\ket{m}$ (i.e., $P_m(t)$) can be expressed as the solution to the master equation $\dot{P_m} = W_{m,n} \times P_m$, where $W_{m,n}$ describes the system-bath interaction induced scattering rate from state $\ket{n}\rightarrow \ket{m}$. Under these approximations, the absorption spectrum can be expressed as a linear superposition of all the energy eigenstates weighted by the oscillator's strengths and broadened by a Lorentzian line shape, \textit{i.e.},
\begin{align}
A_{th}^{\nu_q} = \sum\limits_q f_{p,q} \dfrac{\mid W_{q,q}\mid}{h(\nu - \nu_q) + W_{q,q}^2}
\label{eqn:abs}
\end{align}
where $f_{p,q} = \mid \bra{q} \hat{\mu} \ket{p} \mid^2$ is the oscillator strength for transition $\ket{p} \rightarrow \ket{q}$ with $\ket{p}$ defined as ground state in the {\it cis} or {\it trans} well and state $\ket{q}$ has an energy $\epsilon_q (= h\nu_q) = \bra{q} \hat{H_s} \ket{q}$. We compare the absorption peak frequencies computed using Eq.~(\ref{eqn:abs}) to the experimental values of each isomer\cite{birge_nature_1988}. 

Fluorescence decay of the excited state gives rise to the emission spectra. To account for this, in the master equation $\dot{P_m} = W_{m,n} \times P_m$, the rate kernel corresponding to a photon bath (i.e., $J^{em}(\omega_q) = 4 \pi^2 \omega_q^3/3\epsilon_0 (2\pi c)^3$) is included in the theoretical model with vacuum permittivity $\epsilon_0$ and speed of light $c$ \cite{dodin_coherent_2016}. At a certain frequency $\nu_q$, the magnitude of the generalized rates connecting the states with energy difference $\epsilon_q(=h\nu_q)$ governs the intensity of the spontaneous emission spectra. Under the Bloch-secular approximation, the latter can be expressed as
\begin{align}
g_{th}^{\nu_q} \varpropto \int\limits_0^\infty dt \vert \sum\limits_{m,n} W_{m,n}^{em} P_n(t) \delta\left( \nu_{m,n} + \nu_q \right) \vert \;.
\end{align}
In practice, a window function (with width on the order of the average nearest-neighbor energy gap) can be used instead of $\delta$ function, and following this a rectangular function is employed, that is, $\delta\left( \nu_{m,n} + \nu_q \right) \rightarrow \Pi\left( \nu_{m,n} + \nu_q \right) \Pi\left( -\nu_{m,n} - \nu_q - \Delta \nu \right)$.
When comparing to an experimental emission spectrum corresponding to a given excitation frequency,  we take the photon occupation number to be a square wave function centered at that frequency and a finite width of 10 cm$^{-1}$ and zero everywhere else. The latter assumption is justified since the thermal occupation at optical frequencies can be ignored at room temperature, where the experimental spectra were measured.
We compare the theoretically predicted emission spectra $g_{th}^{\nu_q}$ with the experimental data $g_{ex}^{\nu_q}$ obtained by using continuous wave laser at three different excitation frequencies for rhodopsin sample \cite{kochendoerfer_spontaneous_1996}.

\section{Multi-objective Bayesian optimization}
\begin{figure}[!h]
\begin{center}
\includegraphics[width = 6.5 in]{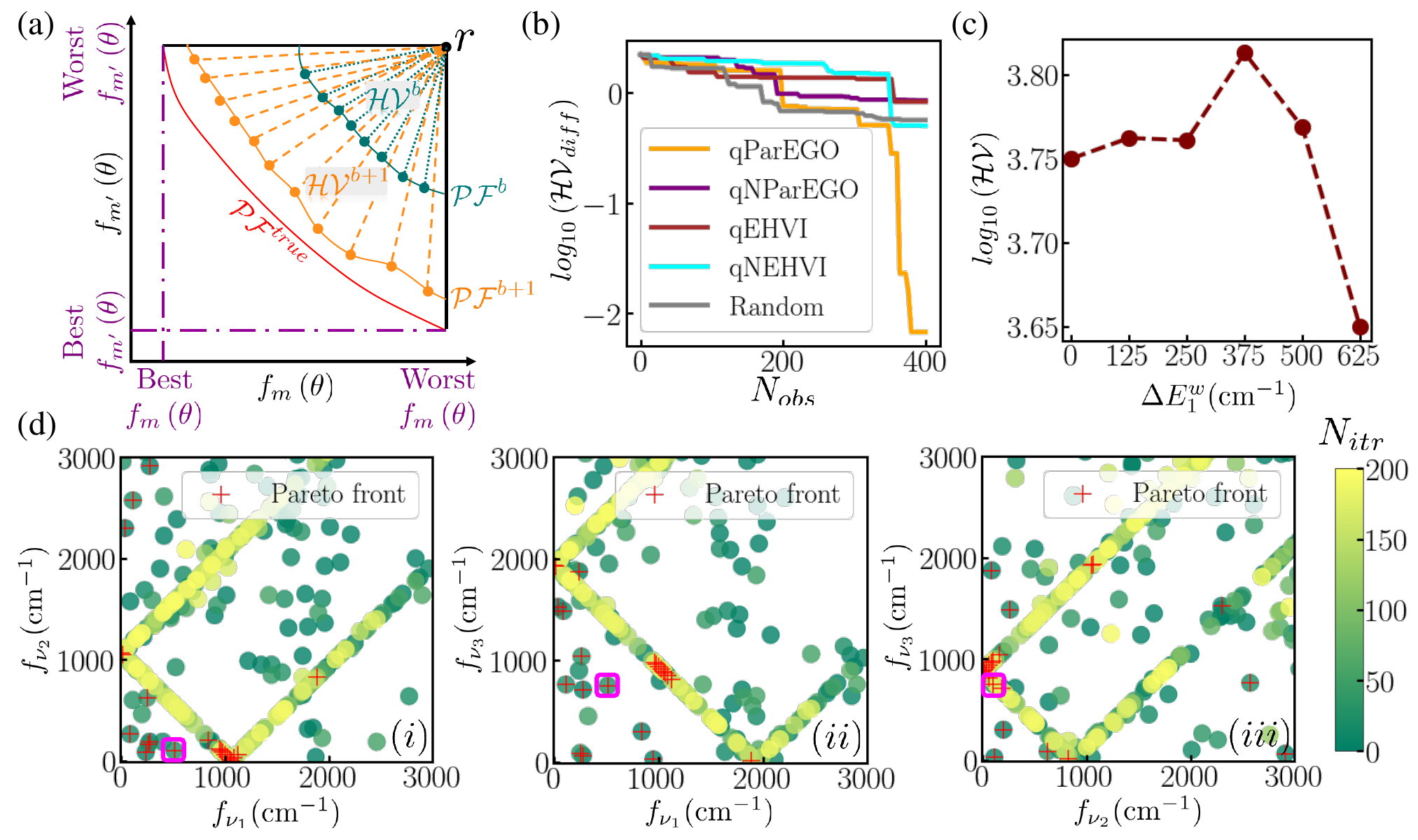}
\caption{(a) An illustration of the iterative improvement of $\mathcal{PF}$ and $\mathcal{HV}$ for two objectives. (b) The efficiency of the four acquisition functions surveyed, represented by the hypervolume $\left( \mathcal{HV}_{diff} \right)$ as functions of the number of observations $N_{obs}$ (beyond initially selected points) for $d = 4$ ($M = 3$) search (objective) space parameters. (c) The hypervolume$\left( \mathcal{HV} \right)$ as a function of the distribution width of $E_1$. (d) The $\mathcal{PF}$ for the $TM$ model with $\Delta E_1^{w}\sim 375$ cm$^{-1}$ by using the constraint $E_1 - W_1 =11211$ cm$^{-1}$($= 1.39$ $eV$). Each panel displays the $\mathcal{PF}$ for different pairs: (i) $f_{\nu_1}$ vs $f_{\nu_2}$, (ii) $f_{\nu_1}$ vs $f_{\nu_3}$ and (iii) $f_{\nu_2}$ vs $f_{\nu_3}$.  The best point (i.e., $\theta^b$) is circled with magenta line . The gradient of the color reflects the number of iterations beyond the initial data points. By using a scrambled sobol sequence, we used randomly sampled 100 initial data points (i.e., $\{\delta\theta_d\}$). A batch size of $Q = 2$ is used for the acquisition function with 200 iterations ($N_{itr}$),\textit{i.e.}  $N_{obs} = N_{itr}\times Q$.}
\label{Pareto_375}
\end{center}
\end{figure}

Multi-objective optimization\cite{deb_multi_2001} is used to optimize the vector-valued 
function 
$F(\theta)=[f_1(\theta),\allowbreak f_2(\theta),\cdots,f_m(\theta),\cdots,f_{M-1}(\theta),f_M(\theta)]$ with $M\geq2$ and $\theta$ as the search space parameter in a compact search space $\Theta$, i.e, $\theta \in \Theta$. The dimension of the search space is $d$ such that $\Theta \subset \mathcal{R}^d$. Each individual $f_m(\theta)$, denoting a specific objective, can be approximated with a multi-variate Gaussian process, \textit{i.e.},
\begin{align}
f_m \sim \mathcal{GP}\left( \mu_m, \mathcal{V}_m \right)\;,
\end{align}
where $\mu_m$ ($\mathcal{V}_m$) represents the mean (covariance) of the objective. By using $N$ training points, \textit{i.e.}, $\{ (\theta^n,f_m^n) \mid n = 1,2,3,\cdots N \}$, in the Gaussian process, the mean value of the new point $\theta^{\star,q}$ and variance can be written as \cite{snoek_practical_2012}
\begin{align}
\mu_m(\theta^{\star,q}) &\simeq \mathcal{K}^\top(\theta^{\star,q},\Xi)[\mathcal{K}(\Xi,\Xi) + \sigma_{noise} \mathcal{I}]^{-1} Y_m \nonumber\\
\mathcal{V}_m(\theta^{\star,q}) &\simeq \mathcal{K}(\theta^{\star,q},\theta^{\star,q}) -  \mathcal{K}^\top(\theta^{\star,q},\Xi)[\mathcal{K}(\Xi,\Xi) + \sigma_{noise} \mathcal{I}]^{-1} \mathcal{K}(\theta^{\star,q},\Xi)
\end{align}
where $\Xi = \{\theta^1, \theta^2,....,\theta^{N-1},\theta^N \}$ and $Y_m = \{ f_m^1, f_m^2,....,f_m^{N-1},f_m^N \}$ contain prior information. The diagonal noise $\sigma_{noise}$ is included in training observations for a better posterior. The connection (or strength) among the data points is established by the covarience matrix $\mathcal{K}(\cdot,\cdot)$, which is estimated by using the Matern $5/2$ ARD kernel function \cite{daulton_differentiable_2020,daulton_parallel_2021}. To select the new search space points, i.e., $\Theta_{new} = \{ \theta^{\star,1}, \theta^{\star,2},..,\theta^{\star,q}, ..., \theta^{\star,Q-1},\theta^{\star,Q} \}(=\{ \theta^{\star,q} \}^Q_{q=1})$, BO uses an acquisition function that balances highly uncertain exploring regions and exploits areas believed to be optimal. Then the true function is evaluated using $\Theta_{new}$ (which makes it computationally less expensive to evaluate the black-box function). 
We tried four different acquisition functions (named qParEGO, qNParEGO, qEHVI, and qNEHVI, see Ref. \cite{daulton_differentiable_2020,daulton_parallel_2021}) to identify the Pareto frontier ($\mathcal{PF}$) [see Fig.\ref{Pareto_375}a]. 
Basically, a vector $F(\theta^\prime)$ pareto dominates $F(\theta^{\prime\prime})$ (i.e., $F(\theta^\prime) < F(\theta^{\prime\prime})$), if $F(\theta^\prime) \leq F(\theta^{\prime\prime})$ and $\exists$ $m$ $\in$ $\{1,2,\cdots,M \}$ such that $F^{(m)}(\theta^\prime) < F^{(m)}(\theta^{\prime\prime})$. Thus, a $\mathcal{PF}$ can be defined over a set of objective vectors $\mathcal{F} = \{F(\theta^\prime) \mid \theta^\prime \in \Xi \subseteq \Theta \}$ as $PARETO(\mathcal{F}) = \{F(\theta^\prime) \in \mathcal{F} : \nexists \theta^{\prime\prime} \in \Xi \quad s.t. \quad F(\theta^{\prime\prime}) < F(\theta^\prime) \}$ and set of all these corresponding $\theta^\prime$ values describe Pareto set of optimal designs that is $\Theta^{opt} = \{ \theta^\prime \}$. However, since it is too difficult to find the true $\mathcal{PF}$ (i.e., $\mathcal{PF}^{true}$ see Fig.\ref{Pareto_375}a), a useful metric described as hypervolume ($\mathcal{HV}$) is commonly used to find the quality of the approximate $\mathcal{PF}$ estimated by Bayesian optimization \cite{yang_multi-objective_2019}. For a set of points $\mathcal{Y} \subset \mathcal{R}^M$, the $\mathcal{HV}$ (i.e., $\mathcal{HV}(\mathcal{Y},r)$) can be defined as the M-dimensional Lebesgue measure $\lambda_M$ of the region dominated by $\mathcal{P}:=PARETO(\mathcal{Y})$ and bounded from above by a reference point $r\in \mathcal{R}^M$ [see Fig.\ref{Pareto_375}a]. It is schematically illustrated in Fig.\ref{Pareto_375}a, where it is assumed that with the increase of iteration from $b$ to $b + 1$, new points explored by an acquisition function improve the surrogate model close to the actual/local maxima or minima. It leads to the iterative improvement of the $\mathcal{PF}$ from $\mathcal{PF}^{b}$ to $\mathcal{PF}^{b+1}$. This improvement in the $\mathcal{PF}$ pushes $\mathcal{PF}^{b+1}$ away from the reference point $r$ compared to the $\mathcal{PF}^{b}$ [see Fig.\ref{Pareto_375}a]. 
It implies an iterative enhancement in the $\mathcal{HV}$ from $\mathcal{HV}^b$ to $\mathcal{HV}^{b+1}$.
In other words, the $\mathcal{HV}$ increases as the $\mathcal{PF}$ improves with iterations and moves away from the reference point. This process continues till a particular number of iterations, beyond which the enhancement in the $\mathcal{HV}$ stops and, therefore, improvement in $\mathcal{PF}$ ceases. 
In this manner, the best $\mathcal{PF}$ could be observed by analyzing $\mathcal{HV}$ for a particular acquisition function.
In our simulation, we tried four different acquisition functions. Now, to compare the performance of different acquisition functions, a new metric termed hypervolume difference ($\mathcal{HV}_{diff}$) is defined such that $\mathcal{HV}_{diff}$ = $\mathcal{HV}_{sel}$ - $\mathcal{HV}_{itr}$, here $\mathcal{HV}_{sel}$ is the value of the $\mathcal{HV}$ selected after applying all acquisition functions (a value little higher than the maximum obtained $\mathcal{HV}$) and $\mathcal{HV}_{itr}$ is the iteratively measured hypervolume for each acquisition function at each iteration. 
Evidently, the decrease of $\mathcal{HV}_{diff}$ implies closeness to the best observed $\mathcal{PF}$. The acquisition function that provides the minima of $\mathcal{HV}_{diff}$ could be used for further exploration and exploitation of the search space.


\section{Pareto Front with original PES}

We compare the results from using four different acquisition functions: qParEGO, qNParEGO, qEHVI, and qNEHVI in addition to iterative random selection of points [see Fig.\ref{Pareto_375}b] to find the $\mathcal{PF}$ and $\mathcal{HV}$ for single trajectory calculations. The reference point used in this case is $r = [2000, 2000, 2000]$. As expected, $\mathcal{HV}_{diff}$ decreases with the number of observations in the case of $d=4,M=3$ [see Fig.\ref{Pareto_375}b]. Moreover qParEGO performs the best in our calculations and is subsequently used to estimate $\mathcal{PF}$ for the ensemble calculations, detailed below. 

We next analyze the effect of ensemble averaging in terms of the width of the Gaussian-distributed $E_1$, $\Delta E_1^{w}$, especially on the quality of the predicted emission spectra using the $\mathcal{HV}$ as the figure of merit. We find that $\mathcal{HV}$ is not a monotonic function of $\Delta E_1^{w}$ [see Fig.\ref{Pareto_375}c]. Specifically, $\mathcal{HV}$ increases as we increase the width from zero (single trajectory) to $\sim 125$ cm$^{-1}$, and shows a small drop at $\sim 250$ cm$^{-1}$. The maximum of $\mathcal{HV}$ is observed at $\sim 375$ cm$^{-1}$, implying the best $\mathcal{PF}$. $\mathcal{HV}$ decreases with further increase in distribution width.  That is, we find the optimal width for the $E_1$ parameter under the current scheme is $\Delta E_1^{w}\sim375$ cm$^{-1}$.  

This nonmonotonic performance in fitting the experiental emission spectra can be rationalized by considering two factors.  First, as justified in the previous section, the shift of the emission peak in tandem with the shift of excitation wavelength can be expected from an inhomogeneously broadened ensemble where each of constituent systems has similar dynamical properties, as is properly descibed using our current approach. For example, when shone a redshifted excitation beam, the part of the ensemble that reacts to the excitation would have a smaller electronic gap to begin with.  Provided similar relaxation dynamics to systems with different optical gaps while holding all other parameters the same,  we expect that the emission peak shift to be concurrent with the shift in the excitation wavelength up to the width of the optical gap ($E_1$) distribution.  Second, with increasing width, such dynamical homogeneity is expected to deteriorate; that the relaxation dynamics of different subensemble selected by different excitation conditions becomes dissimilar.  The optimality of the optical gap distribution can thus be understood as a balance between these two factors.  As such, we expect this optimality in optical gap distribution to hold for more complicated and realistic models as well, which is left for future investigation.


In the following we focus on the $\mathcal{PF}$ of the optimal width $\Delta E_1^{w}\sim375$ cm$^{-1}$.  The $\mathcal{PF}$ is shown in Fig.\ref{Pareto_375}d.  We find a reasonably good agreement with the experimental data with points close to the lower left corner of $\mathcal{PF}$. The three error functions (i.e., $f_{\nu_m}$) can not be simultaneously minimized to $\sim 0$ cm$^{-1}$, however, by using Bayesian optimization we successfully observed a data point $\theta^b$ (summarized in Table 1 
and highlighted in magenta in Fig.\ref{Pareto_375}d), with which all the predicted peak positions (of ensemble averaged trajectories) were found close to target, \textit{i.e.} $f_{\nu_m}\leq 650$ cm$^{-1}$. Note that, this minimal error is significantly small compared to the selected wide range (i.e., $\sim 12000$ cm$^{-1}$) used to estimate the peak positions of emission spectra. Additionally, the best data point $\theta^b$ is chosen based on the lowest sum of all the error functions.


\section{Pareto Front with modified PES}

\begin{figure}[!h]
\begin{center}
\includegraphics[width = 6.5 in]{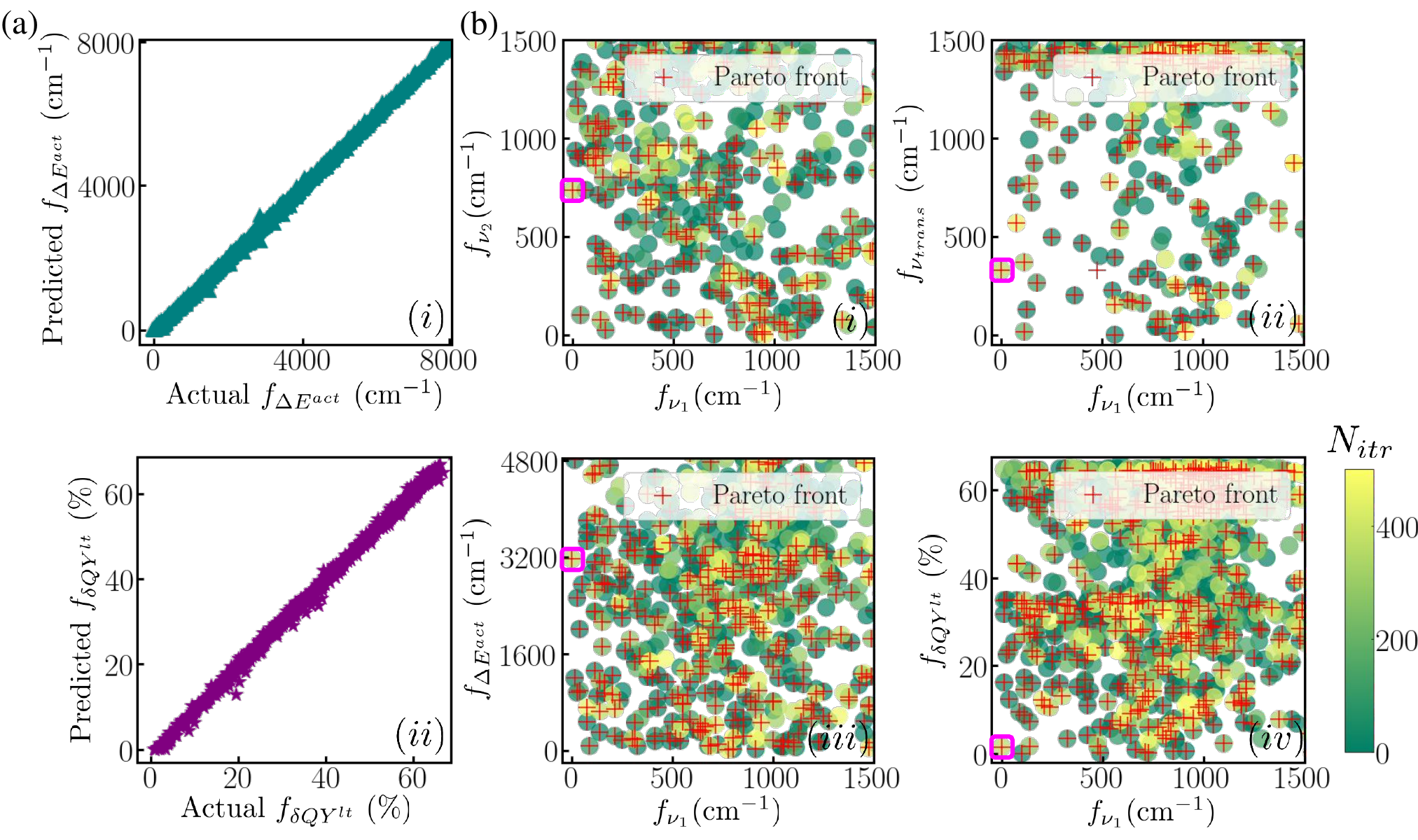}
\caption{(a) Illustration of the model training for ($\rNum{1}$) $f_{\Delta E^{act}}$ and ($\rNum{2}$) $f_{\delta QY^{lt}}$. (b) The $\mathcal{PF}$ for the $TM$ model by using single trajectory without the constraint $E_1 - W_1 =\Delta E_{ex}$ for $d = 7$ ($M = 8$) search (objective) space parameters. Each panel displays the $\mathcal{PF}$ for different pairs: ($\rNum{1}$) $f_{\nu_1}$ vs $f_{\nu_2}$, ($\rNum{2}$) $f_{\nu_1}$ vs $f_{\nu_{trans}}$, ($\rNum{3}$) $f_{\nu_1}$ vs $f_{\Delta E^{act}}$ and ($\rNum{4}$) $f_{\nu_1}$ vs $f_{\delta QY^{lt}}$.  The best point (i.e., $\theta^b_{add}$) is circled with magenta line. The gradient of the color reflects the number of iterations beyond the initial data points. By using a scrambled sobol sequence, we used randomly sampled 500 initial data points (i.e., $\{\delta\theta_d\}$). A batch size of $Q = 2$ is used for the acquisition function with 500 iterations. 
}
\label{Pareto_obj8}
\end{center}
\end{figure}


By incorporating all eight objectives in the target space, we are able to achieve a high-quality model training, as demonstrated in Fig.\ref{Pareto_obj8}a. Using these well-trained models, Bayesian optimization is employed to identify Pareto points that minimize the error functions. In the case of single trajectory, the $\mathcal{PF}$ exhibits a specific point (highlighted in magenta on Fig.\ref{Pareto_obj8}b) that allows for simultaneous minimization of all the targets. Again, the ensemble averages centered around this specific point are evaluated. Even for this Pareto point, an optimal ensemble width is observed to illustrate accurate $cis$ absorption.


\bibliography{Paper_Rhodopsin_Bayesian}

\end{document}